# Distributed simulation of city inundation by coupled surface and subsurface porous flow for urban flood decision support system

V.V. Krzhizhanovskaya [a,b]*, N.B. Melnikova [a,b], A.M. Chirkin [b], S.V. Ivanov [b], A.V. Boukhanovsky [b], P.M.A. Sloot [a,b]

[a] *University of Amsterdam, The Netherlands*
[b] *National Research University ITMO, St. Petersburg, Russia*

**Abstract**

We present a decision support system for flood early warning and disaster management. It includes the models for data-driven meteorological predictions, for simulation of atmospheric pressure, wind, long sea waves and seiches; a module for optimization of flood barrier gates operation; models for stability assessment of levees and embankments, for simulation of city inundation dynamics and citizens evacuation scenarios. The novelty of this paper is a coupled distributed simulation of surface and subsurface flows that can predict inundation of low-lying inland zones far from the submerged waterfront areas, as observed in St. Petersburg city during the floods. All the models are wrapped as software services in the CLAVIRE platform for urgent computing, which provides workflow management and resource orchestration.

## 1. Introduction.

Floods are the most common and frequent natural disasters. Flood-induced economic losses are shockingly huge! Just one flood brought by the hurricane Katrina cost 125-250 billion US dollar damage [1], [2], which accounted for 5-10% of the total US federal budget in 2005. Averaged over the past 30 years, floods around the world killed 6,753 people per year and claimed an annual economic loss of 13.7 billion USD [3], out of which 4 billion USD in Europe [4]. Most of these damages are impinging on the cities and urbanized areas. The average number of affected people was 96,878,672 per year, a population of 2 countries like Spain or 6 countries like the Netherlands.

---

\* Corresponding author. *E-mail address:* Valeria.Krzhizhanovskaya@gmail.com.

In some cases, it is possible to prevent floods by monitoring flood defense conditions and by detecting the weak spots early enough to repair or reinforce the levee. In other cases, preventing floods is impossible, but we can mitigate the consequences and save human lives by alarming them and by suggesting the best evacuation routes. In the past decade, many international projects have been developing flood early warning systems and disaster management decision support systems, see [5] and references therein. They do not replace the engineering work on building and reinforcing flood protection systems, but they give us the time and the means to monitor the situation and to react promptly.

Two recent projects exemplify the progress in this field. One is the *UrbanFlood* European project [6] that developed a flood early warning system, combining the developments in monitoring dikes with sensor techniques [7], [8], physical study of dike failure mechanisms [9], software for dike stability analysis [10], [11], simulation of dike breaching, flood, and city evacuation [12], [13], [14]. All the data streams, the models and the computational resources have been connected via the Common Information Space, an advanced ICT infrastructure [15].

The second example, presented in this paper, is the flood management decision support system developed in the Advanced Computing Lab of the ITMO University, St. Petersburg, Russia [16]. It combines the meteorological predictions with modeling the storm winds, the Baltic Sea water levels, the long waves and seiches (standing waves) in the Gulf of Finland. This information is then used to optimize the operation of flood barrier gates [17], [18], [19] and to calculate possible scenarios of city inundation and citizens' evacuation [14]. The models are integrated into the CLAVIRE (CLoud Applications VIRtual Environment) platform that orchestrates the modeling workflow and provides computational resource management [17].

In the next section, we give some background information on floods in St. Petersburg and the state-of-the-art in coupling the models of surface (overland) flow with subsurface flow that we address in this paper. The remainder of the paper is organized as follows: Section 3 describes the architecture of our flood management decision support system; Sections 4 and 5 present the overland and subsurface flow models respectively; implementation details are given in Section 6; the first simulation results of coupled inundation models are described in Section 7; and Section 8 completes the paper with conclusions and future plans.

## 2. Floods in St. Petersburg and the state-of-the-art in coupling models of surface and subsurface flows

St. Petersburg is a 5-million-population city, the former capital of the Russian Empire; it accumulated a grand cultural and architectural heritage and very expensive businesses and industries. And all that treasure lies in the lowland of the Neva River delta, with the historical center lying at the sea level or mere 1-4m above it. Over 300 large-scale floods have been recorded in the city history [20]. Figure 1 shows the location of St. Petersburg and flood statistics for the past 3 centuries.

One of the new challenges we faced in developing a decision support system for flood management in St. Petersburg was the subsurface water flow, previously unaccounted for in the city inundation models. Even during a minor flood in St. Petersburg, water is often observed emerging from manholes and underground passages far away from the flooded waterfront. This is caused by the water that enters through the stormwater drain inlets and other holes in the inundated area; driven by the pressure head, it flows through the drain pipes and other natural or man-made hollow spaces around the underground pipes (e.g. gas, sewage and communication systems); and comes out of the open holes and drain inlets, especially in the lower grounds. Figure 2 illustrates the problem.

Why is it important to model the subsurface water flow that only re-distributes the flood? - Because it changes the rules in evacuation planning, since some inland streets can be closed shortly after the inundation started at the waterfront areas. In the early projects that tried to address this problem, the overland inundation models were enriched with the boundary conditions on the land surface that took into account an estimated balance of the surcharge and discharge of the manholes and drain sewers [21]-[23]. Recent progress in computational methods and computing power allowed to simulate also the subsurface flows and to couple these models to the overland flows [24]-[25]. Some researchers suggested "extended" shallow water equations, which are suitable for both surface and subsurface flows [26]-[27], however only 1D simulations have been performed so far. Keeping in mind the requirement of faster-than-real-time simulations in critical situations, fully coupled surface-subsurface 3D simulations are still feasible only for small catchments, not for a city scale. We therefore adopted a distributed simulation approach, where the two models run on different servers and exchange information on the land surface boundary.

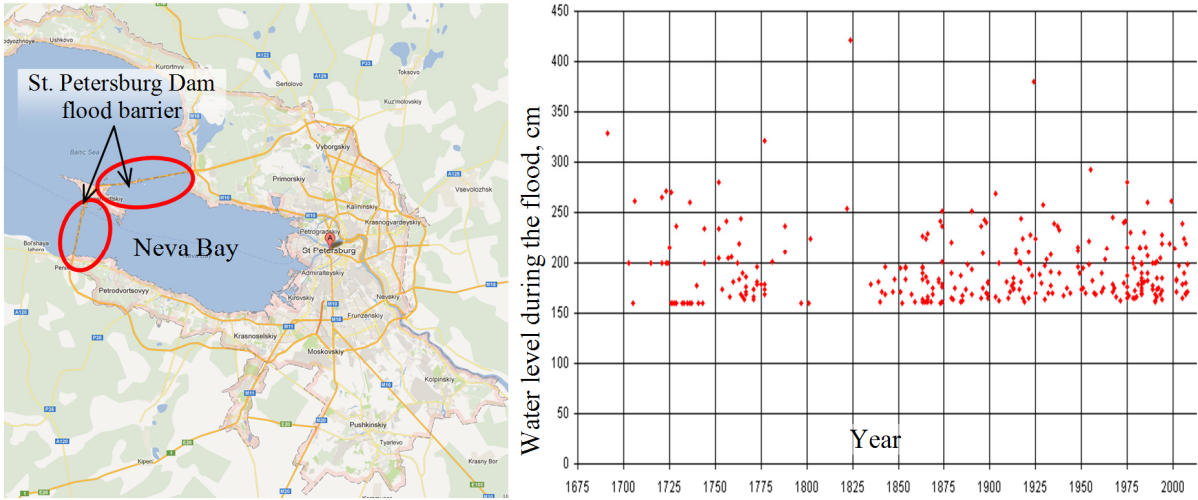

Fig. 1. <u>Left</u>: St. Petersburg city is located in the lowland delta of the Neva River (image from Google Maps). A 25-km flood barrier dam protects the city since its completion in 2011. <u>Right</u>: Over 300 floods have been recorded in the city history in the past 3 centuries. The graph shows water levels in cm, each point represents one flood event [20].

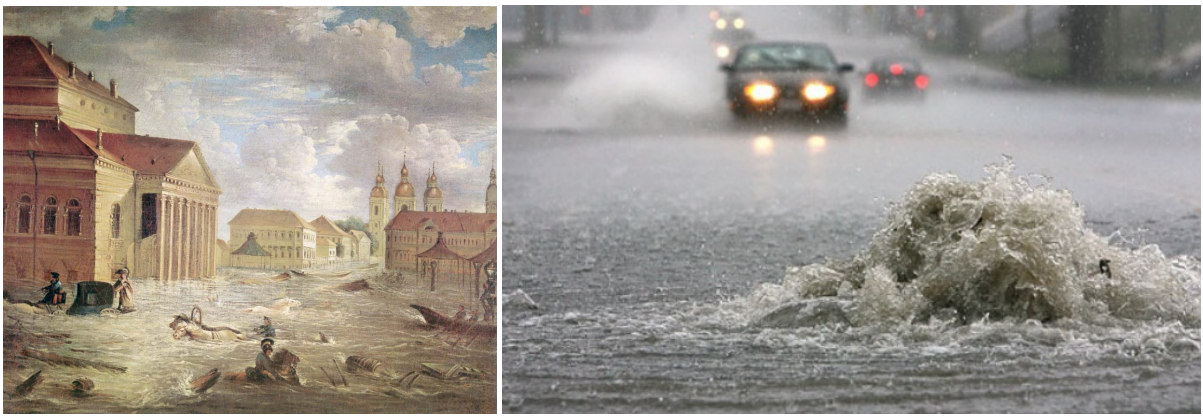

Fig. 2. <u>Left</u>: The worst ever flood in St. Petersburg on the 19th of November 1824. Painting by Fedor Alekseev, who died 4 days after that disastrous flood. <u>Right</u>: Storm water jets out of a manhole. Photo by Todd Yates, Corpus Christi Photos.

## 3. Urban flood decision support system and modeling workflow

Our flood management decision support and early warning systems include a number of computational models and data assimilation procedures. Figure 3 (left) shows the modeling workflow. First, a **data assimilation** module collects and analyses historical and present data on sea levels and winds correlated to the weather conditions; based on that, **meteorological predictions** are made for the next 48 hours, using the models HIRLAM, NCEP/NCAR, ECMWF. These predictions together with the current observations are used for modelling the **atmospheric pressure and wind** fields by the SWAN model. Then the BSM-2010 and FMI models calculate the **sea levels** in the Baltic Sea and the long waves and seiches (standing waves) in the Gulf of Finland. It gives a range of **water levels in the reference point** located near Saint Petersburg State Mining Institute (Gorny Institute).

If predicted water levels exceed the flood threshold of 160 cm, St. Petersburg Dam flood barrier shall close the gates. A special module **optimizes the gate operation plan**. This is not so trivial because in addition to the storm surge waters coming from the sea, the Neva River brings about 2500 m$^3$ per second, or 9 million ton per hour. If the dam gates are closed, this immense amount of water will accumulate in the small Neva Bay (see Fig. 1, left) and raise water levels by 2.5 cm per hour. The main goal of this optimization is to close the gates for the shortest possible period of time that still provides safe water levels in the city. This part of the workflow has been implemented in a **Decision support system for flood barrier gate operation** (highlighted as shown in Fig. 3, left), and described in detail in [17] and [28]. In addition, another model has been developed that simulates **the under-gate flows** in order to avoid potentially dangerous vibration modes, seabed scour (erosion) around the gates, and some other aspects of gate operation (ecological, economical, technical maintenance, etc.), see [18], [19]. Integration of this model into the gate decision support system is in progress.

If the **flood is possible** (in spite of the optimal floodgates operation or in case of gates failure), the **Flood**

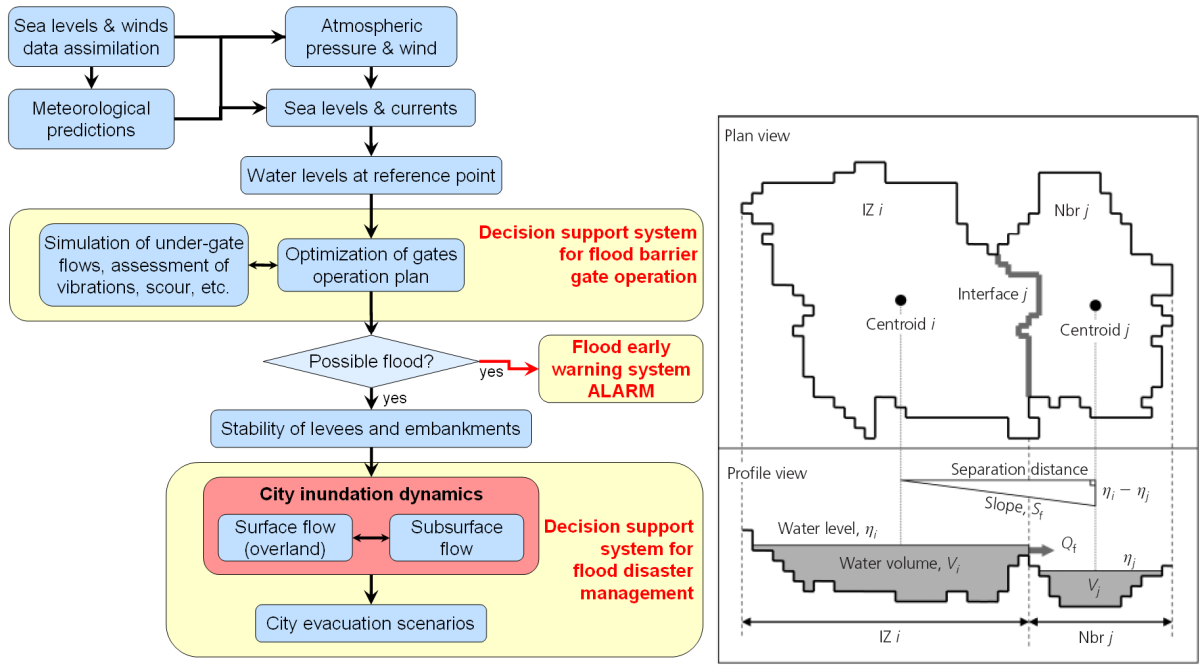

Fig. 3. <u>Left</u>: Modeling workflow for flood early warning and decision support systems. <u>Right</u>: Schematic of an Impact Zone (IZ) with a neighbour, in plan and profile. Solid grey represents a volume of water [29].

**early warning system** issues an ALARM and starts additional models: one module can calculate the **stability of levees and embankments**, based on the Virtual Dike model [11]. Another module simulates the **city inundation dynamics**, given the predicted water levels. This simulation consists of two models: **Surface (overland) flow** and **Subsurface flow**. They exchange information on the land surface boundary. In the next sections we describe these models and show some preliminary results. Finally, given the flood dynamics, we can calculate **city evacuation** or emergency rescue scenarios with the last model in the workflow. The results of city inundation and evacuation provide information to the **Decision Support System for flood disaster management**. The models are glued via the CLAVIRE workflow and resource management platform, and presented to the users via an interactive graphical user interface on a multi-touch screen or via a simplified web interface.

## 4. City inundation model: surface (overland) flow

In critical situations decisions must be made very fast, therefore the requirement on all our simulations was a response within minutes, with a maximum of a few hours for long-term predictions. This restriction determined the methods and codes selected for simulation of inundation on a city scale. The surface flow dynamics is simulated by the Dynamic Rapid Flood Spreading Model (DRFSM), developed by the HR Wallingford team and adapted in our project. The model is based on a computationally efficient volume spreading approach, sufficiently robust for use in flood risk models [29]. The limitations of this model are that it is not suitable for very fast flooding processes like tsunamis, and it can provide only indirect means for assessment of building damages due to the water flows, since it does not solve for the energy balance equation.

In a pre-processing stage, the domain is discretised in irregular shaped computational elements. These so-called Impact Zones (IZs) are delineated around depressions in the topography (se Fig. 3, right). Input to this pre-process is the floodplain topography in the form of a Digital Terrain Model (DTM). Each IZ captures the underlying topography by the means of a table giving the volume of water stored in the IZ for different flood levels. This mesh allows to speed up the simulation by reducing the number of computational elements compared to the initial number of cells in the input DTM. The use of the level-volume relation means that the computation of the water level in an IZ is more precise than the use of an averaged ground level for situations where the IZ is not entirely flooded.

The model receives flood volumes discharged into floodplain areas from breached or overtopped defenses and then spreads the water over the floodplain according to the terrain topography. Spreading of flood water is achieved by transferring water between IZs at each computational time-step. The discharge between IZs can be calculated by two methods, the Manning relationship (i.e. similar to diffusion wave models) or the weir relation. The computational time-step is constant. Water level, average discharge and average velocity are calculated in each IZ during the computation. The water level is then used as a boundary condition in the subsurface flow model. To take into account water surcharge due to the subsurface flow or water sink through the storm-water drain inlets, each inundation zone can be assigned the rate of water level change in m/s. Water surcharge is dynamically calculated by the subsurface flow model.

This approach can be used in probabilistic flood risk analysis where multiple runs are required, or in real time situations (flood forecasting), where the model run time is critical. The model has been validated against available flood data and more advanced models, see [29] and references therein.

## 5. Subsurface flow model (flow through porous media)

The subsurface model we developed assumes that the city network of storm-water drainage pipes, drain inlets and outlets is sufficiently dense; therefore on a scale of the city, we can consider the land being a porous medium characterized by some porosity and permeability. These parameter values are different from those of the real soils (sands or clays), because they represent the density and "throughout" of the underground water streams. As a test case, we consider Vasilyevsky Island of St. Petersburg city that faces the Gulf of Finland and hosts a historical city centre district with over 200,000-population.

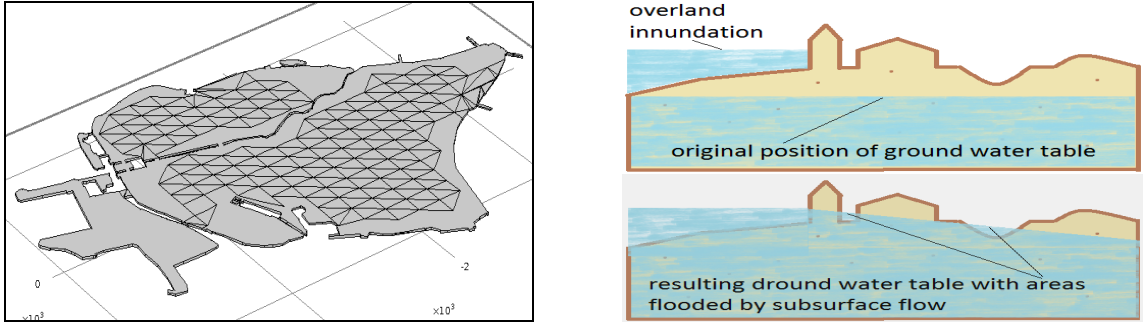

Fig. 4. Left: Vasilyevsky Island simulation domain. Right: Coupling of the overland and subsurface flow models. Top: initial position of the ground water table and overland inundation. Bottom: flood resulting from the subsurface flow model.

A 3-dimensional simulation domain is presented in Figure 4 (left). It spans 5 km by 7.5 km in horizontal direction and 20 m in vertical direction. The upper surface of the island and vertical surfaces forming the embankments are treated as water-permeable. The water levels from surface flow model are used in the subsurface flow model as boundary conditions, see Fig. 4 (right).

The resistance of the porous medium to the water flow is modeled using a transient groundwater flow equation based on Darcy's law [30]:

$$S\frac{\partial p}{\partial t} + \nabla \cdot \left[-\frac{K_S}{\mu}\nabla(p + \rho g z)\right] = 0 \qquad (1)$$

Here $S$ is water storage [1/Pa]; $p$ is water pressure [Pa]; $t$ is time [s]; $K_S$ is permeability [m$^2$]; $\mu$ is water dynamic viscosity [Pa·s]; $g$ is standard gravity [m/s$^2$], $\rho$ is water density [kg/m$^3$], $z$ is land elevation coordinate [m]. Filtration velocity is calculated from Darcy's law: $\vec{V} = K_s/\mu(\rho\vec{g} - \nabla p)$

Equation (1) is solved with the boundary conditions and with the initial condition specified as follows: Boundary condition on the bottom surface of the island simulates an impermeable wall: $V_n = 0$. Pressure boundary condition is specified on the embankments and on the upper surface of the island: $p = \rho g \cdot (h_{in}(t,x,y) - z)$, where $h_{in}(t,x,y)$ is the transient water level above ground obtained from the surface (overland) flow model; $x$ and $y$ are coordinates in a Cartesian coordinate system. In "dry" areas, a zero water head is specified: $h_{in}(t,x,y) = 0$. The initial condition assumes that the ground water table stays at the reference sea level: $p = -\rho g z$. The surcharge water levels resulting from the porous flow simulation are computed from the pressure values on the upper surface of the island:

$$h_{filtr}(t,x,y) = \frac{p(t,x,y)}{\rho g} \text{ for } p(t,x,y) \geq 0; \quad h_{filtr}(t,x,y) = 0 \text{ for } p(t,x,y) < 0$$

Negative pressure values on the upper surface indicate that phreatic surface is located below the ground surface in this area and no inundation occurs due to the subsurface filtration. The total floodwater depth is calculated as a sum of the levels obtained from the surface (overland) flow model and from the filtration analysis: $h_{total}(t,x,y) = h_{in} + h_{filtr}$.

## 6. Implementation of city inundation models and integration in CLAVIRE distributed environment

The surface (overland) flow model is implemented in C/C++, the auxiliary codes for fields' interpolation and setting boundary conditions and model parameters are written in C# and Python. All input parameters and simulation results are stored in a database. Minimal hardware requirements are: Intel x86 Family 6 processor, 1 GB free disk space and 1 GB RAM memory. Minimal software requirements include Windows OS with .NET 2.0 and Microsoft SQL Server 2008. The simulation executable and the database are installed in a virtual machine (VM). The VM image is deployed under Xen hypervisor. A typical model for city inundation contained 500,000 computational cells pre-processed into 100,000 Impact Zones. For a given city topography and impact zone combination, the computational runtime can take from seconds to minutes per time step, depending on the number of zones where the inundation starts and on the amount of discharged water.

The subsurface model is implemented in Comsol Multiphysics® 4.3 using the finite element method and time dependent implicit BDF solver. Each time step, a system of nonlinear algebraic equations is solved by Newton's iterative method with a parallel MUMPS (MUltifrontal Massively Parallel sparse direct Solver). Fields interpolation and integration of the module in the computational workflow is implemented by the scripts written in Matlab language with the help of LiveLink™ for Matlab® component of the Comsol package. The scripts implement automatic execution of the following operations:
- export of inundation water levels from the surface flow model to the subsurface flow model;
- start of subsurface flow model in a batch mode from CLAVIRE environment;
- export of subsurface simulation results into CLAVIRE;
- visualization of floodwater depths (time-step images and animations of flood dynamics).

The subsurface simulation can run on a hardware platform with at least Pentium IV or Athlon processor, 10 GB free disk space and 1 GB RAM, with the minimal requirements on software being: Windows XP2, or Linux 2.6.18 with GNU C Library 2.3.6, or MacOS 10.5. The visualization software requires x86 1.6 GHz processor with SSE2 support, 1 GB RAM, and a videocard 512 MB and DirectX10. The software requirements are Windows XP SP 4 with .NET 4 and Microsoft XNA Framework 4.0. A typical subsurface simulation with 50,000 computational cells requires 1 GB memory and takes 10 seconds of CPU time for 30 minutes of physical time on a quad-core 2.4 GHz processor in a multithreading mode.

The overall decision support modeling workflow is implemented in CLAVIRE (CLoud Applications VIRtual Environment) platform described in [31]. It creates and maintains the data links between the models in a distributed environment and orchestrates the computational resources with the goal of satisfying the urgent computing requests. All the software components are wrapped as services using domain-specific language EasyFlow and EasyPackage. The CLAVIRE environment supports the urgent computing paradigm by a scheduling mechanism with different priority levels and with a feature of matching simulation software with available computational resources, following the concept suggested in [32], [33]. The main idea is that a special module in CLAVIRE environment runs a series of benchmarks and measures simulation time with different combinations of input parameters, on different hardware platforms and with different number of parallel cores/nodes. It analyses this information and makes the best software-hardware match given the requirements on simulation deadline [34], [35]. All new simulation runs are adding data to the benchmark statistics, thus improving the forecast and helping to deliver the results in time.

## 7. Results

Since its completion in 2011, St. Petersburg Dam flood barrier gates were closed several times to prevent city inundation. Figure 5 (left) shows an example of the schedule for gate closing and opening moments suggested by the decision support system for flood barrier gates operation. The black dotted line shows water level dynamics at the tip of Vasilyevsky Island predicted by the BSM-2012 model in case of open flood barrier gates. Two solid lines show two acceptable scenarios of gate operation that prevent flood. No flood was expected if the gates were closed according to the suggested plans, therefore the lower half of the modeling workflow (Fig. 3, left) was not processed. To test the coupled surface-subsurface city inundation model, we assumed that the gates stay open and that the inundation threshold is 130 cm (blue dashed line in Fig. 5, left). The excess water level (above 130 cm) was assumed to start overtopping all the waterfront areas simultaneously. In reality, there is a delay of several minutes, but that can be neglected in the first tests. The discharge through the boundary element was calculated by the weir equation, knowing the length of the boundary and the water head.

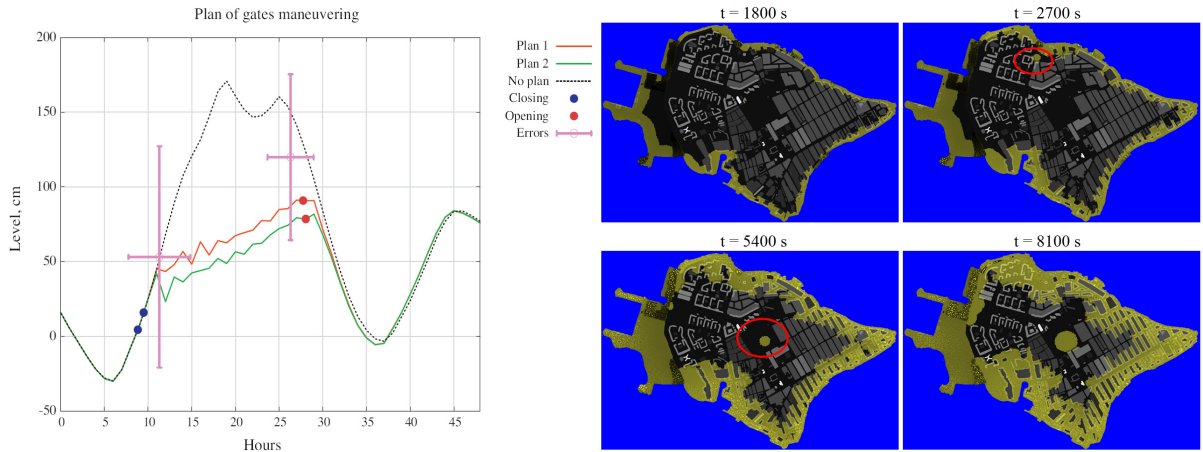

Fig. 5. <u>Left</u>: Water levels for three scenarios: flood barrier gates open (dashed line), and two scenarios of gates closing and opening moments suggested by the decision support system for flood barrier gates operation [17].
<u>Right</u>: Visualized results of inundation simulation. The elevation map was artificially modified by lowering the central part of the island, to illustrate the effect of subsurface flow, which emerged in the concave area in the center.

An example of visualized results of flood simulations is shown in Fig. 5 (right). At moment t = 1800 s, the inundation is seen along the perimeter of the island. At moment t = 2700 s, we see a small area that was flooded further inland from the waterfront area affected by the overland water flow. No other puddles were initially observed. To illustrate the effect of the subsurface flow, the elevation map of Vasilyevsky Island was artificially modified by lowering the central part of the island. It produced an additional inundated area in the center at t = 5400 s, which grew in size, as we see at t = 8100 s. In the final version of the paper, we plan to present some performance results of the models and of the CLAVIRE environment. In the conference, we will demonstrate also the results of evacuation modeling coupled to flood simulations [14].

## 8. Conclusions and future plans

To prevent floods and to mitigate their consequences, we developed a decision support system for flood early warning and disaster management. It includes the models for data-driven meteorological predictions, for simulation of atmospheric pressure, wind, long sea waves and seiches; a module for optimization of flood barrier gates operation; models for stability assessment of levees and embankments, for simulation of city inundation dynamics and citizens evacuation. The system has been tested on St. Petersburg city that faces the Gulf of Finland and has a long history of severe floods. Recently completed St. Petersburg flood barrier now protects the city, but requires a smart decision support system for flood barrier gate control. The modelling workflow successfully predicted the floods and proposed a schedule for closing and opening the flood gates.

The novelty of this paper is a coupled distributed simulation of surface and subsurface flows that can predict inundation of low-lying inland zones far from the submerged waterfront areas, as observed in St. Petersburg city during the floods. Preliminary results of inundation simulations on an artificially modified topography of Vasilyevsky Island with a lowered part of the island showed that the models correctly simulate the effect of subsurface water flow, which emerged in the concave inland area. In the future, the coupled modeling shall be tested on a realistic topography different from a flat pancake.

All the models have been wrapped as software services in the CLAVIRE platform for urgent computing, which provides workflow management and resource orchestration. One of the features important for urgent computing is smart scheduling and matching simulation software with the available computational resources, with the goal of delivering the results in time for making an informed decision in critical situations. We are currently working on it and on other resource management aspects. In the final version of the paper, we plan to present some performance results of the models and of the CLAVIRE environment.

In the modeling workflow, we plan to connect the model of under-gate flow simulations to the module optimizing the flood barrier gate operation. This model has been developed and tested separately, now it needs to be plugged in, with some strategy for prioritizing the various aspects of flood decisions support: the probability of flood, the potentially dangerous flow modes causing gate vibrations, the undesirable seabed erosion (scour) around the gates, and some negative ecological and economical aspects. Next, the evacuation model shall be connected to the CLAVIRE environment. It was running manually so far. Then we will look into modeling pollution from the stormwater runoff, and into an automated damage assessment.

**Acknowledgements**

This work was supported by the EU FP7 project UrbanFlood, grant N 248767; by the Leading Scientist Program of the Russian Federation, contracts 11.G34.31.0019 and 13.G25.31.0029; and by the BiG Grid project BG-020-10, # 2010/01550/NCF with financial support from the Netherlands Organisation for Scientific Research NWO. We also thank the HR Wallingford team, especially Julien L'Homme, Ben Gouldby, Jonathan Simm and Andrew Tagg, for providing a free license for the DRFSM software used in this project and for their help with simulation of a free-surface flow for city inundation.